\documentclass[11pt]{article}
\usepackage[xdvi]{graphicx}
\usepackage{epsfig,axodraw,cite}
\usepackage{amsfonts, amssymb, amsmath}
\usepackage{subfigure}

\newcommand{\be}{\begin{equation}}
\newcommand{\ee}{\end{equation}}
\newcommand{\ba}{\begin{eqnarray}}
\newcommand{\ea}{\end{eqnarray}}
\newcommand{\ban}{\begin{eqnarray*}}
\newcommand{\ean}{\end{eqnarray*}}
\newcommand{\bml}{\begin{mathletters}}
\newcommand{\eml}{\end{mathletters}}

\def\quarter{\textstyle{1\over4}}

\def\ie{{\it i.e. }}
\def\eg{{\it e.g. }}

\def\L{{\cal{L}}}

\def\k{{\cal{K}}}

\newcommand{\nn}{\nonumber}

\def\de{\partial}
\def\a{\alpha}
\def\b{\beta}
\def\g{\gamma}
\def\G{\Gamma}
\def\d{\delta}
\def\D{\Delta}
\def\e{\eta}

\def\La{\Lambda}
\def\k{\kappa}
\def\m{\mu}
\def\n{\nu}
\def\r{\rho}

\def\p{\pi}

\def\ep{\epsilon}
\def\th{\theta}
\def\Th{\Theta}
\def\z{\zeta}

\def\nn{\nonumber}

\begin{document}

\begin{titlepage}

\title{Self-properties of codimension-2 braneworlds}

\author{{\bf Christos Charmousis}$^{1,*}$ and {\bf Antonios
Papazoglou}$^{2,**}$ \vspace{5mm}\\
 $\phantom{1}^1${\it LPT, Universit\'e de Paris-Sud,}\\
{\it B\^at. 210, 91405 Orsay CEDEX, France}\\
\\$\phantom{1}^2${\it Institute of Cosmology and Gravitation,}
\\{\it University of Portsmouth, Portsmouth PO1 2EG, UK}\\}

\date{\today}

\maketitle

\begin{abstract}

We consider four-dimensional de Sitter, flat and anti de Sitter
branes embedded in a six-dimensional bulk spacetime whose dynamics
is dictated by Lovelock theory. We find, applying a generalised
version of Birkhoff's theorem, that all possible maximally
symmetric braneworld solutions are embedded in Wick-rotated black
hole spacetimes of Lovelock theory. These are warped solitonic
spaces, where the horizons of the black hole geometries correspond
to the possible positions of codimension-2 branes. The horizon
temperature is related via conical singularities to the tension or
vacuum energy of the branes. We classify the braneworld solutions
for certain combinations of bulk parameters, according  to their
induced curvature, their vacuum energy and their effective
compactness in the extra dimensions. The bulk Lovelock theory
gives an induced gravity term on the brane, which, we argue,
generates four-dimensional gravity up to some distance scale.
 As a result,  some simple solutions,
such as the Lovelock
corrected Schwarzschild black hole in six dimensions, are shown to
give rise to self-accelerating braneworlds. We also find that
several other solutions have self-tuning properties. Finally, we
present regular gravitational instantons of Lovelock gravity and
comment on their significance.

\vspace{4mm}


\noindent \textbf{Report No\/}: LPT08-37

\vspace{.3cm}

\begin{flushleft}

$^{*}~$ e-mail address: christos.charmousis@th.u-psud.fr \\
$ ^{**}$ e-mail address: antonios.papazoglou@port.ac.uk

\end{flushleft}

\end{abstract}

\end{titlepage}

\section{Introduction}

Combined cosmological  and astrophysical data indicate that at
least $96\%$ of the actual matter content of the Universe has yet
to be detected in particle accelerators. This is a correct
statement provided  we assume a homogeneous universe described by
Einstein's field equations. One fourth of this yet unseen
component, dark matter, we can hope to discover in the Large
Hardron Collider and is an expected, quite natural, cosmological
signature of particle physics theories such as low energy
supersymmetry (SUSY). For the remaining chunk, usually dubbed dark
energy, that has cropped up more recently from supernovae data
\cite{super}, the situation is a lot less clearer. Data  is best
"fitted" by assuming a tiny positive cosmological constant of
magnitude $(10^{-3}{\rm eV})^4$. The mass scale
associated to this energy is that of, roughly, what one expects
for neutrino masses, a really tiny number in the cosmological
arena. This cosmological constant is so small because the radius of the observed
universe is by now huge, of  size of the square inverse to the
mass scale in question. It is dominant because dark matter is
extremely diluted at cosmological scales of roughly $10$Gpc whereas
 super clusters of galaxies such as Virgo are at $20$Mpc.

Unlike dark matter, from the point of view of theoretical particle
physics, the cosmological constant is rather problematic. For the
"natural" value of the cosmological constant, from the point of
view of Quantum Field Theory, is of the order of the ultraviolet
cutoff we would impose for our quantum field theory, ranging from
the Planck scale to the SUSY breaking scale, depending on our
theoretical prejudice. The fine tuning we impose to compensate
this vacuum energy by some bare, gravitational cosmological
constant is the biggest known discrepancy between theoretical
expectations and our understanding of experimental data. This is
the cosmological constant problem \cite{cc} that was there already
there \cite{weinberg} when the precision of existing data pointed
towards a value, assumed for simplicity to be zero.

By admitting the presence of a small, but not null, cosmological
constant we add two additional problems to the "old" cosmological
constant problem \cite{weinberg}, namely, why is it so small and not zero, and why
is its density now of the same order as dark matter density. Why
therefore can we observe it now, \ie the right time that it is
possible for us to observe it in the first place? Dark energy
models usually accompanied by scalar fields with suitable
potentials do not solve this fine-tuning problem. They are also
generically problematic due to enhanced radiative corrections
associated to the scalar(s) \cite{lorenzo} coupling to matter.
Although general relativity is very well tested at solar system
scales, both at weak and strong field regimes \cite{will}, it is
not so at cosmological scales. In fact, to get an idea of the
colossal difference in scales, take the quotient of the Hubble
radius over the Astronomical unit (earth to sun distance) to get
$10^{15}$! Therefore, cosmological data, if we do not
assume some extra source of unknown matter or cosmological constant, tell us
precisely that gravity is modified at the infrared.

Taking into account the above discussion, modifying gravity in the
infrared  is a legitimate theoretical hypothesis that should be
taken seriously. As yet, no convincing modification of general
relativity has been found. Often the problem is experimental,
constraints as for example for $f(R)$ theories in the solar system
and in galaxy clusters \cite{luca1}. However, there are also
theoretical difficulties since a large distance modification
spoils our understanding of asymptotic behaviour, which is very
well-defined in General Relativity (GR). Typically, such models
\cite{longdistance,dgp}  have ghost instabilities
\cite{specter,kal,kal2} and strong coupling problems \cite{dvali2}
which are as yet not understood and therefore are no better than
fitting data with a single, however small parameter, the
cosmological constant.

This of course does not mean there is no such consistent
modification. Clearly one has to try more complex setups and some interesting
toy models \cite{stealth} have been proposed and are now being put
to the test. By far, however, the most successful and popular
modification is that of self-acceleration \cite{deffayet} present
in the Dvali-Gabadagze-Porrati (DGP) model \cite{dgp}. There, the
small current acceleration of the Universe is understood as a
geometric effect originating from the (unfair) competition between
the five-dimensional and four-dimensional curvature scales.
Unfair, because the resulting length scale, the crossover scale,
is of the order of the Hubble horizon today, $H_0\sim 10^{-34}
eV$. At these length scales gravity becomes five-dimensional with
a very low five-dimensional gravitational scale, roughly of $10^4
{\rm GeV}$ and we enter a geometric acceleration phase.
Unfortunately, this solution has a ghost in its linear
perturbation spectrum \cite{kal,kal2}. Furthermore, the results
are fuzzed by the presence of strong coupling problems
\cite{dvali2} and either the theory loses linear predictability
(which makes it useless cosmologically) or is completely unstable
(see \cite{koyama} for exact solutions  and beyond linear order).
To summarize, therefore, self-acceleration teaches us that we can
in principle explain in a simple way the current acceleration of
the Universe by a geometric effect of a codimension-1 braneworld,
but the geometry modification required to attain this result is
not seemingly viable. Along these lines, recently \cite{tolley}
have given an idea on similar terms where the cosmological
constant results from a higher order gravity correction in a
codimension-1 braneworld.

An even more ambitious proposal that emerged in braneworld models
is that of the self-tuning and has to do with the "old"
cosmological constant problem \cite{weinberg}. The idea was to
find models where the tension of the brane, that is its vacuum
energy, can be large without affecting the curvature of the brane
and without fine-tuning of it with other brane or bulk parameters
\cite{5dself}. The best attempt to realize such a kind of model
has been in six dimensions and in the case that the brane is of
codimension-2 \cite{CLP}. These conical branes have the special
property that their tension merely induces a deficit angle in the
surrounding bulk geometry and does not curve the brane
world-volume. The brane curvature is induced from the bulk
dynamics and is not directly related to the brane vacuum
energy{\footnote{We emphasize here that we will not necessarily
consider flat brane curvature, but also cases where a small brane
curvature is allowed with a large vacuum energy on the brane.}}.
Several models with codimension-2 branes were studied with various
compactifications \cite{6dself,6dsigma}, however most of them had
hidden fine-tunings or curvature singularities present. Although
self-tuning vacua can be found, where indeed solutions of the same
curvature correspond to  a (continuous) range of brane vacuum
energy, a cosmological constant problem resolution would require a
dynamical selection mechanism of such vacua. This is because, as a
rule, there exist nearby solutions of different curvature, which
introduce the question whether  the self-tuning solution is an
attractor.

In the present paper, we will examine a  completely novel
possibility of obtaining acceleration due to geometry as well as
certain self-tuning properties. The modified gravity theory that
we will study is Lovelock theory \cite{Lovelockth} in six
dimensions, which is the natural extension of GR in higher
dimensions. The Lovelock theory in six dimensions has in addition
to the Einstein-Hilbert term the Gauss-Bonnet combination.
Although the latter is a topological invariant in four dimensions,
it becomes dynamical for higher dimensions and modifies the
gravitational theory. The codimension-2 branes present in the
generic vacua of this theory, can have interesting properties,
similar to the ones stated in the previous paragraphs. We will
find for example self-accelerating cases with non-compact internal
spaces as well as self-tuning vacua for several compact and
non-compact vacua. These examples open new possibilities for
consistent self-acceleration and effective self-tuning which need
to be considered in more detail in the future.

In fact, Lovelock theory plays a crucial double role, not only in
the bulk, but also for the brane junction conditions
\cite{bostock}. On the one hand,
 it provides geometric self-acceleration and novel solutions which are absent in Einstein theory,
and on the other hand, it gives  an induced gravity term on the
brane \cite{bostock,zcc},  much like in DGP \cite{dgp}. Here we
must emphasise, that this does not mean that we expect to have
 Einstein gravity on the codimension-2 braneworld, rather, it means
that up to some scale, gravity "looks" four-dimensional as in the
DGP model.  Lovelock theory has the extraordinary geometric
property to induce an Einstein-Hilbert term at the level of the
junction conditions \cite{zcc}.

The structure of the paper is as followss: we will firstly present
the general black hole solutions in six-dimensional Lovelock
theory and then show how by double Wick-rotation we can obtain the
most general axially symmetric braneworlds with maximally
symmetric codimension-2  branes. Then we will scan through several
classes of solutions and analyse the cases of interest, \ie the
self-accelerating and the self-tuning vacua introducing each time,
as few "extra bulk parameters" as possible. We will briefly
comment on the physical implications of some new regular instanton
solutions of Lovelock gravity and we will finally conclude.

\section{Static black hole solutions} \label{secBH}

Let us consider the six-dimensional dynamics of gravity
 with a bare cosmological constant $\La$ and a Gauss-Bonnet (GB)
term (see Appendix A for a more general system with in addition a
gauge field coupled to gravity). The action of the system reads
\be \label{chaaction} S=\int d^6 x\, \sqrt{-g}\left[\frac{1}{16 \pi
G_6}(R +\hat{\alpha} \L_{GB})-2\La\right]  ~, \ee where
\be\L_{GB}=R_{MNK \La}R^{MNK \La} -4R_{MN}R^{MN}+R^2 ~, \ee is the
Gauss-Bonnet Langrangian density, $G_6$ the six-dimensional
Newton's constant and $\hat{\a}$ the Gauss-Bonnet coupling.

It is straightforward to write down the Einstein equations of
motion for the above action. They read \be G_{MN} -
\hat{\a}H_{MN}=-8 \pi G_6 \La  g_{MN}~, \label{papein}\ee with
$G_{MN}=R_{MN} - \frac{1}{2}R g_{MN}$ the Einstein tensor and the
following Gauss-Bonnet contribution
 \ba H_{MN}&=&\frac{1}{2}\L_{GB}~g_{MN}-2RR_{MN}+4R_{MK}R_N^{~K} \nn \\&&+4R_{KM\La
N}R^{K\La}-2R_{MK \La P} R_M^{~~K \La P}~. \label{papH}  \ea

Our ultimate goal is to find the maximally symmetric solutions of
the above equations of motion. For this purpose, let us consider a
four-dimensional space of maximal symmetry parametrised by
$\kappa=0,-1,1$, with metric \be h_{\m\n} = {\d_{\m\n} \over
\left(1+{\k \over 4} \d_{\m\n}x^\m x^\n \right)^2}~. \label{paph}
\ee These spaces are four-dimensional flat space, hyperboloid or
sphere respectively of curvature $R[h]=12\k$. A generalised
version of Birkhoff's theorem for (\ref{chaaction}) states,
\cite{birk} (see \cite{zegers} for the Lovelock version) that
every  six-dimensional spacetime solution of (\ref{papein}) having
such four-dimensional maximal sub-spaces (\ref{paph}) is locally
isometric to \be \label{chabh} ds^2=-V(r)
dt^2+\frac{dr^2}{V(r)}+r^2 h_{\m\n} dx^\m dx^\n~, \ee admitting
therefore a locally timelike Killing vector ($\partial_t$ in the
coordinates of (\ref{chabh})). The solution of the potential $V$
for the equations of motion (\ref{papein}) is found to be
\cite{boulware,cai} \be \label{chapot} V(r)=\kappa +
\frac{r^2}{2\alpha}\left[1+\epsilon \sqrt{1+4\alpha
\left(a^2-\frac{\ep \mu}{r^{5}}\right)}\right]~, \ee where the
parameters appearing in the action (\ref{chaaction}) have been
rescaled to, \ba \label{chapar1} \alpha=6\hat{\alpha}~,\qquad
16 \pi G_6 \Lambda=20a^2>0&~&\mbox{(for}~ dS_6)~, \qquad \\
a^2=-k^2<0&~&\mbox{(for}~ AdS_6)~. \nn \ea

The integration constant $\m$ is \be \label{chapar2}
\mu=\frac{4\pi G_6 M}{\Sigma_{\kappa}}~, \ee where  $M$ is the AD
or ADM mass of the solution and $\Sigma_{\kappa}$ is the area of
the unit four-dimensional maximally symmetric subspace. Finally,
$\epsilon=\pm 1$, giving rise to two distinct branches of
solutions. The convention of the $\mu$ sign is chosen so that the
gravitational mass is always $\mu>0$. Indeed, as one can easily
check by expanding the square root for large distances, the sign
flip in front of $\mu$ is necessary to match the Schwarzschild de
Sitter solution behaviour for positive AD mass. The case where
$1+4\a a^2=0$ is special, because the theory can be written in a
Born-Infeld (BI) form \cite{zanelli}.

Setting $\mu=0$ gives us the asymptotic vacua of the theory
(\ref{chaaction}) which unlike Einstein  theory are not unique
(for given bare parameters in (\ref{chaaction})). Then, we find
that asymptotically we have an effective six-dimensional
cosmological constant \be
 16 \pi G_6 \Lambda_{eff}=-\frac{10}{ \alpha}\left[ 1 + \ep \sqrt{1+4\a a^2} \right]~, \label{papL}
\ee with a normalisation like in (\ref{chaaction}). We note in
particular, that $\alpha$ gives an effective cosmological constant
in the bulk even without a cosmological constant term in the
action, \ie even with $a^2=0$. The maximally symmetric space that
the solutions asymptote to, depends on the sign of $\a$ and the
branch of the solution, \ie $\ep$. It is interesting to expand the
above expression at $\a \to 0$ in order to check if there is an
Einstein theory limit for the above vacua or not \be 16 \pi G_6
\Lambda_{eff}=-\frac{10}{ \alpha}(1+\ep) -20 \ep a^2  + {\cal
O}(\a) ~. \label{antexpL}\ee

Hence, we distinguish the following three cases

\begin{itemize}

\item  For $\epsilon=+1$, the solution, as can be seen by
(\ref{antexpL}), does not have an Einstein theory limit as $\a
\rightarrow 0$, although there may be a relevant $dS_6$ or $AdS_6$
Einstein solution mimicking (\ref{chabh}). We name this branch the
{\it Gauss-Bonnet branch}. The solution asymptotes $AdS_6$ space
for $\alpha>0$ and to $dS_6$ space for $\alpha<0$.  Unlike what is
argued in \cite{boulware} this branch is not, at least,
classically unstable for the reasons put forward by
\cite{chadestek} (for a full study of this see \cite{padilla}).

\item For $\epsilon=-1$, the solution has an Einstein theory limit
(as $\a \rightarrow 0$) as seen from (\ref{antexpL}) and therefore
we call this branch the {\it Einstein branch}. The solution
asymptotes $AdS_6$ space for $\a>0$, $-1/(4\a) < a^2<0$ or $\a<0$,
$a^2<0$, to $dS_6$ space for $\a<0$, $0 < a^2<1/(4|\a|)$ or
$\a>0$, $a^2>0$ and to $M_6$ for $a^2=0$.

\item For the special BI case $1+4\a a^2=0$, as we can see from
(\ref{papL}) we obtain an $dS_6$ or $AdS_6$ asymptotic solution
which does not have an Einstein theory limit and no possible flat
vacuum. However this is the only case that we have  a unique
vacuum.

\end{itemize}

Let us now proceed into analysing  the solution at hand
(\ref{chabh}) (we will mostly follow \cite{ms}). For the above
metric solution (\ref{chabh}), there are two possible
singularities in the curvature tensor, the usual $r=0$, and also a
branch cut singularity at the (maximal) zero of the square root
\be \label{chasing} r_s^5={4 \a \ep \mu \over 1+4\alpha a^2}~. \ee
Whenever $r_s$ is real and positive, this is the singular end of
spacetime (\ref{chabh}). For the BI case $1+4\alpha a^2=0$ there
is no such singularity.

We will have a black hole solution if and only if there exists
$r=r_h$ such that $V(r_h)=0$ and $r_h>r_{max}$, where
$r_{max}=\max\{0,r_s\}$. Indeed the usual Kruskal extension \be
dv_{\pm}=dt\pm \frac{dr}{V(r)}~, \ee gives that $(v_+,r)$ and
$(v_-,r)$ constitute a regular chart across the future and past
horizons of (\ref{chabh}). It is actually straightforward and
rather useful to show the following: $r=r_h$  is an horizon iff,
\ba &&r_h>r_{max}~, \label{papcon1} \\ &&\epsilon(2\alpha \kappa+r_h^2)\leq 0~, \label{papcon2}\\
&&p_\a(r_h)=0 ~{\rm with} ~ p_\alpha(x)=-a^2 x^5+\kappa x^3
+\alpha \kappa^2 x +\ep \mu~. \label{papcon3}  \ea In addition,
one should make sure that the sign of $V$ on each side of the
roots of the above polynomial is such that (\ref{chabh}) describes
a black hole. The sign of $p_\a$ and $V$ are not in general the
same, depending on the signs of $\ep$ and $\a$.

Therefore for $\epsilon=+1$  we have $r_h^2\leq -2\alpha \kappa$,
\ie, the event horizon is bounded from above. This means that for
$\kappa=0$ there are no black hole solutions in this branch. Also
note that $p_{\alpha=0}(x)$ is just the usual polynomial for a
Einstein black hole, and hence, $\alpha$ couples only to the
horizon curvature $\kappa$. Therefore, for $\epsilon=-1$ and
$\kappa=0$ the horizon positions are the same as in the GR
solutions. In fact, when $\epsilon=-1$ we have similar structure and
properties of the solutions (\ref{chabh}) as their Einstein counterparts.

\section{Codimension-2 braneworlds}

A standard procedure for Einstein theory to generate brane world
solutions from black hole solutions \cite{wu,muko} is to perform a
double Wick rotation for a black hole solution. The same will be
valid for (\ref{chaaction}) and the black hole solution
(\ref{chabh}) at hand. We set $t \rightarrow i \theta$ and in
addition make a further Wick rotation  $x^0 \to i t$ in the metric
$h_{\m\n}$ (\ref{paph}) so that it becomes of Lorentzian signature
\be h_{\m\n} = {\e_{\m\n} \over \left(1+{\k \over 4} \e_{\m\n}x^\m
x^\n  \right)^2}~. \ee
 Then, the maximally symmetric spacetime sections
correspond to four-dimensional Minkowski, $AdS_4$ and $dS_4$ for
$\k=0,-1,1$ respectively with curvature $R[h]=12\k$. The solutions
are now of manifest axial symmetry with $\partial_\theta$ as the
angular Killing vector, \be \label{chasol} ds^2=V(r)
d\theta^2+\frac{dr^2}{V(r)}+r^2 h_{\m\n}dx^\m dx^\n~. \ee  The
six-dimensional spacetime has the correct symmetries to describe a
maximally symmetric four-dimensional brane world. The staticity
theorem invoked for black hole spacetimes \cite{birk} tells us now
that axial symmetry comes for free and need not be imposed for
resolving the system of equations. The solutions (\ref{chasol})
are therefore the general solutions describing maximally symmetric
branes, something that has been overlooked up to now even in the
case of Einstein theory. It is interesting to note that in the
case of Einstein theory in four dimensions, we obtain in this way
the general maximally symmetric gravitating cosmic strings, which
can differ drastically from their flat counterparts (see in
particular, the cosmic string solutions in $AdS$ presented in
\cite{linet}).

The positions of the horizons $r_h$ will be the endpoints of the
internal space and the solutions will have meaning if we keep the
spacelike regions of the previous black hole solutions, \ie the
ones  with $V(r)>0$. Let us suppose that we keep the space-like
region between two horizons $r_-$ and $r_+$, with $r_- < r_+$ [the
subsequent discussion applies of course also in the case that we
have only one horizon and we keep the side which is spacelike]. At
these endpoints of spacetime, which are also the fixed points of
the axial symmetry, one can in general put branes of dimension
four, in other words 3-branes. Since the brane solutions we have
found are maximally symmetric, the brane can only carry some
two-dimensional Dirac charge, \ie pure tension. Taking $x^\mu={\rm
const.}$ and expanding around the zeros of $V$ we get, \be
ds^2\approx \left(\quarter V_{r_\pm}'^2\right)
\rho^2_{\pm}d\theta^2+ d\rho_\pm^2~, \ee with the Gaussian Normal
radial coordinate \be
\rho_\pm=\sqrt{\frac{4(r-r_\pm)}{V'_{r_\pm}}} ~, \ee which  is
well defined in all cases with $V'_{r_\pm} \neq 0$. The case of
$V'_{r_\pm} = 0$ needs special attention as we will see later
(Sec.\ref{Nariai} and Sec.\ref{BInariai}). In this coordinate
system, the brane energy momentum tensor is
$T_{\mu\nu}^{brane}=S_{\mu\nu} \delta^{(2)}=S_{\mu\nu}
\frac{\delta(\rho_\pm)}{2\pi \rho_\pm}$, with  $S_\m^\n= -T_\pm
 \d_\m^\n$, where $T_\pm$ are the brane tensions. If the angular
coordinate has periodicity $\th \in [0,2\pi c)$, then the deficit
angles which are induced at the two brane positions are $\d_\pm =
2 \pi (1 - \b_\pm)$ with \be \b_\pm = {1 \over 2}|V'_\pm|c ~.
\label{antc}\ee The angular periodicity $c$ is an arbitrary
topological integration constant of the solution and can be varied
to generate physically different brane world solutions. Let us
note here, that in a compact model, we can always use the freedom
of choosing $c$ to set the deficit angles of one of the branes to
zero, \eg $\b_-=1$, thus, obtaining a "teardrop" compact space
with only one brane of $\b_+ \neq 1$.

From the Einstein equations (\ref{papein}) supplemented by the
brane tension terms, one can separate the distributional Dirac
parts and write down induced Einstein equations for the branes.
These brane junction conditions are \cite{bostock,zcc} \be 2\pi
(1-\beta_\pm)\left(-\g_{\m \n} +4\hat{\alpha} G_{\m \n} ^{ind}
\right)=8 \pi G_6 S_{\m \n}~, \label{antInduced}\ee where
$\g^\pm_{\m\n}= r_\pm^2 h_{\m\n}$ is the induced metric on the
branes with curvature  $R[\g^\pm]=12\k/r^2_\pm \equiv 12\k
H^2_\pm$, and $G_{\m \n} ^{ind}=-3\k H_{\pm}^2 \gamma_{\mu\nu}$ is
the induced Einstein tensor. It is important to emphasize here
that $r_{\pm}^2$ depends on the geometric parameters of the bulk
solution (\ref{chasol}) namely the mass $\mu$, and the bare
parameters $\alpha$ and $a^2$. It is also effectively the warp
factor of the brane. The induced Newton's constant on the two
branes can be determined from (\ref{antInduced}) to be \be
G^\pm_4={3G_6 \over 4\pi \a(1-\b_\pm)}~. \label{antNewt}\ee Note
that in order to have positive induced Newton's constant, we
should have have angle deficit ($\b_\pm<1$) for $\a>0$ and angle
excess ($\b_\pm>1$) for $\a<0$. The two dimensional warped space
in the coordinates $\theta$ and $r$ has in the latter case the
shape of a pumpkin whereas in the former case the shape of a
football (rugby ball to be more precise). The important
information coming from (\ref{antInduced}) is that the parameters
$\alpha$ and $\beta$ invoke the relation and the possible
hierarchy between the bulk and four-dimensional scales.

Substituting the $G_{\m \n} ^{ind}$ back in (\ref{antInduced}) we
find a relation between the  Hubble parameters on the branes and
the action parameters \be 2\pi
(1-\beta_\pm)\left(\frac{1}{2\alpha}+ \k H_{\pm}^2\right)=\frac{4
\pi G_6}{\alpha} T^{\pm}~, \label{antJunct}\ee which can be
further simplified, if we substitute the deficit angle from
(\ref{antNewt}) and solve for $H^2_\pm$ \be \k H_\pm^2 = -{1 \over
2 \a} + {8\pi G_4^\pm \over 3} T^\pm~. \label{antcontrib}\ee The
above equation is very important and relates  the curvature on the
brane $H_\pm^2$ to its sources, namely the brane tension and the
Gauss-Bonnet coupling. We see that the junction conditions tell us
that the effective expansion $H_{\pm}$ is in one part due to the
Gauss-Bonnet induced cosmological term and in another  part due to
the vacuum energy of the brane. Since $H_\pm=1/r_\pm$, we remind
the reader that the Hubble parameter can be expressed as a
function of parameters appearing in the soliton potential
(\ref{chapot}). Then, according to the specific explicit solution,
the Hubble parameter is constrained and the above relation can
give bounds on the two previously mentioned sources of curvature.

A final comment before discussing the various brane world
solutions has to do with the induced  effective Newton's constant $G^\pm_4$ appearing
 in (\ref{antNewt}).
In a usual Kaluza-Klein dimensional reduction, the effective
Newton's constant is obtained after  substitution of the
graviton's zero mode wavefunction in the action and the
integration of the extra dimensions. This integration, actually,
defines a relation between the effective Newton's constant on the
brane, the higher dimensional Newton's constant and the volume of
the internal space (see Appendix C for the volume calculation for
the general brane world models that we will discuss). However,
because of the presence of the branes and the Gauss-Bonnet bulk
dynamics, the graviton dynamics, as perceived on the brane, have
peculiarities and four dimensional dynamics with an induced
Newton's constant $G_4^\pm$ can be operative, even for internal
spaces of infinite volume \cite{dgp}. This is the interpretation
one should give to the effective $G^\pm_4$ appearing in
(\ref{antNewt}) which can be different from the Newton's constant
obtained via the volume calculation in Appendix C.  For a finite
volume element the actual Newton's constant perceived by the brane
observer  depends on the scale on which we probe gravity on the
brane. This is analogous to what happens for the case of the DGP
model  \cite{dgp} when we embed it in a Randall-Sundrum setup.

\section{Braneworld solutions} \label{secBW}

In this section, we will analyze some particular cases of
brane-world solutions, keeping only some of the parameters in the
potential (\ref{chapot}) each time, which give some interesting
brane world examples. In each subcase, we will present the
important features introducing the fewest parameters possible.
What will interest us in particular are the zeros of $V$  which
will correspond to possible brane locations. We will continue to
refer to roots of $V$ as horizons although this term is strictly
speaking correct only for the black hole solutions (\ref{chabh}).

\subsection{Zero bare cosmological constant}

Let us first choose zero bare cosmological constant $a^2=0$. Then
in all cases of this class,   the horizon position is given by the
solutions of the algebraic equation \be x^3+A x + M =0 ~. \ee The
roots of the above equation are analyzed in Appendix B. In the
present section, we will in addition  apply the constraints
(\ref{papcon1})-(\ref{papcon3}) and the requirement to be in a
spacelike region with $V>0$, in order to find all the possible
vacua.

Firstly, it is evident that for $a^2=0$  there are no $\k=0$ flat
vacua since the polynomial (\ref{papcon3}) has no
roots\footnote{This is true for solutions of the specific ansatz
(\ref{chasol}), which does not include the unwarped flat case. As
we will see in Sec.\ref{flat}, one unwarped flat vacuum for
$a^2=0$  exists.}. The other cases depend on the branch and the
four-dimensional curvature $\k$.

\begin{figure}[t]
\begin{center}
\epsfig{file=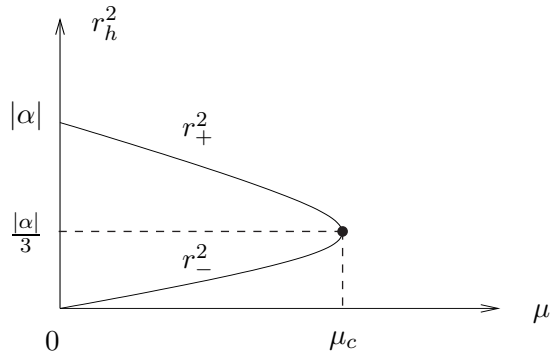,width=6cm,height=4cm}
\begin{picture}(50,50)(0,0)
\Text(-155,110)[c]{$r_h^2$} \Text(10,0)[c]{$\m$}
\DashLine(-172,31)(-65,31){3} \DashLine(-65,31)(-65,3){3}
\Vertex(-65,31){2} \Text(-120,20)[c]{$r_-^2$}
\Text(-120,70)[c]{$r_+^2$} \Text(-65,-10)[c]{$\m_c$}
\Text(-175,-10)[c]{$0$} \Text(-185,31)[c]{${|\a| \over 3}$}
\Text(-185,75)[c]{$|\a|$}
\end{picture}

\end{center}
\caption{The horizon positions $r_-$ and $r_+$ as a function of
the black hole mass $\m$ for the Gauss-Bonnet branch $\ep=+1$ and
for $\a<0$, where the vacua are $dS_4$. For $\m=\m_c$ we obtain
the degenerate horizon case.} \label{2Horizons}
\end{figure}

\subsubsection{The $dS_4$ vacua}

Vacua with $\k=+1$, which are $dS_4$, exist both in the
Gauss-Bonnet and the Einstein branches. As we will see, the
Gauss-Bonnet branch vacua are compact with respect to the $(r,\theta)$ sections, while the Einstein branch
vacua are  non-compact.

\begin{figure}[t]
\begin{center}
\epsfig{file=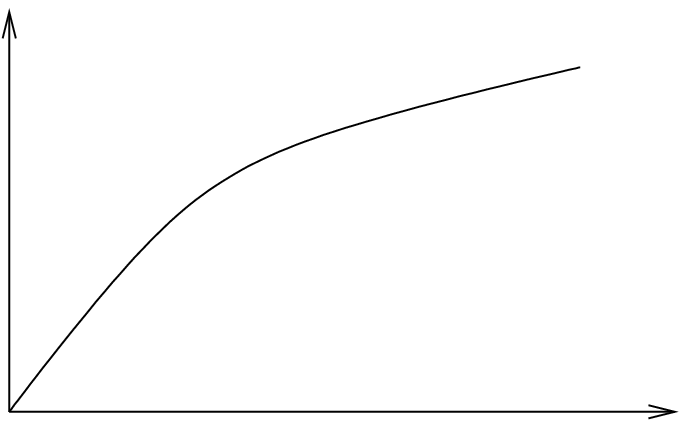,width=5cm,height=3.5cm} \hspace{.5cm}
\epsfig{file=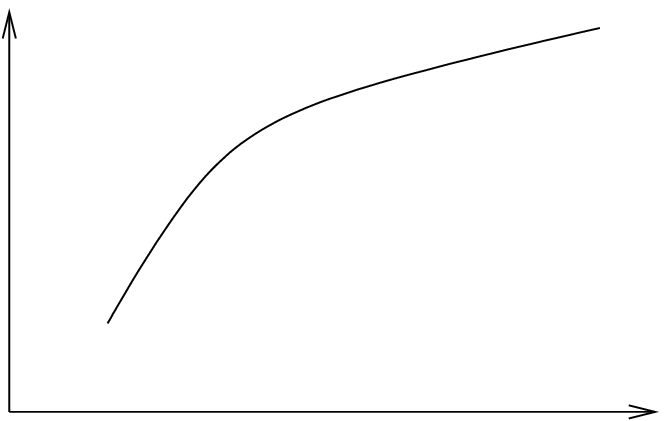,width=5cm,height=3.5cm}
\begin{picture}(50,50)(0,0)
\Text(-290,100)[c]{$r_h^2$} \Text(-125,100)[c]{$r_h^2$}
\Text(0,10)[c]{$\m$} \Text(-165,10)[c]{$\m$}
\DashLine(-142,23)(-122,23){3} \DashLine(-122,23)(-122,3){3}
\BCirc(-122,23){2}  \Text(-122,-10)[c]{$\m_s$}
\Text(-148,-8)[c]{$0$} \Text(-312,-8)[c]{$0$}
 \Text(-155,25)[c]{$2|\a|$}
 \Text(-220,40)[c]{$\a>0$}
\Text(-60,40)[c]{$\a<0$}
\end{picture}
\vspace{-.5cm}

\end{center}
\caption{The horizon position $r_h$ as a function of the black
hole mass $\m$ for the Einstein branch $\ep=-1$, where the vacua
are $dS_4$. The point $\m=\m_s$ corresponds to a singularity $r_s$
and is excluded.} \label{1Horizon-a}
\end{figure}

For $\ep=+1$, the only solution satisfying all the above
requirements is for $\a<0$ and $0 < \m < \m_c \equiv 2
|\a|^{3/2}/(3 \sqrt{3})$. Then we have a double root structure,
which means that the corresponding brane world solution is
of compact $(r,\theta)$ sections. It is important to note that in this case the horizon
positions are bounded from above $r_- \leq \sqrt{|\a|/3}$ and $
\sqrt{|\a|/3} \leq r_{+} \leq \sqrt{|\a|}$, so the corresponding
Hubble parameters $H_\pm=1/r_\pm$ are going to be {\it{bounded
from below}}. When in particular $\m=\m_c$, the two roots become
degenerate and $V'_{r_\pm}=0$. The solution deserves special
attention and will be discussed later on.  The plot of the horizon
(or brane) positions as a function of $\m$ is given in Fig.\ref{2Horizons}.
It is important to stress that the relevant black hole solution corresponding  to this soliton, has the characteristics,
of a  de-Sitter-Schwarzschild like black hole studied in \cite{ms}. Note that the bigger the $\alpha$ the smaller the effective cosmological constant. Here we have  a pure
Gauss-Bonnet soliton (or black hole) in the sense that the $\alpha\rightarrow 0$
limit is singular.

For $\ep=-1$, \ie in the Einstein branch, we have two classes of
solutions. The first class is for $\a>0$ and $\m>0$. The single
horizon of that case can take all positive values. The second
class is for $\a<0$ and $\m> \m_s \equiv \sqrt{2} |\a|^{3/2}$. In
this case the horizon position is bounded from below as  $r_h >
\sqrt{2|\a|}$. Furthermore, in the latter case the singularity
(\ref{chasing}) exists and is hidden behind the horizon. In the
limit $\m \to \m_s$, we have $r_h \to r_s$ and the brane position
becomes singular. We should therefore exclude the $\m=\m_s$ point
from the physical parameter space. In both cases, the brane Hubble
parameter can be as small as desired.
 The plots of the horizon position as a function of $\m$ in the
above cases is given in Fig.\ref{1Horizon-a}. It is interesting to
note that both cases here correspond to a Gauss-Bonnet corrected
six dimensional Schwarzschild black hole solution with $\alpha>0$
or $\alpha<0$.

\begin{figure}[t]
\begin{center}
\epsfig{file=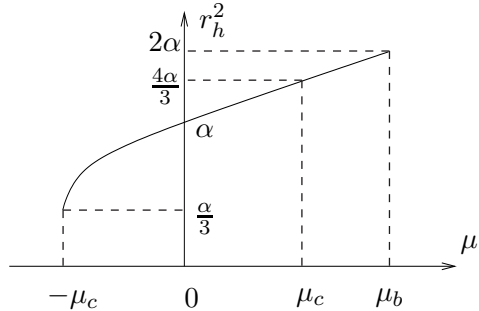,width=6cm,height=3.5cm}
\begin{picture}(50,50)(0,0)
\Text(-97,95)[c]{$r_h^2$} \Text(0,10)[c]{$\m$}
 \DashLine(-106,83)(-30,83){3} \DashLine(-30,83)(-30,2){3}
 \DashLine(-106,72)(-63,72){3} \DashLine(-63,72)(-63,2){3}
 \DashLine(-153,2)(-153,23){3} \DashLine(-153,23)(-108,23){3}
 \Text(-115,70)[c]{${4 \a \over 3}$}\Text(-115,87)[c]{$2 \a $}
\Text(-100,53)[c]{$\a$} \Text(-100,20)[c]{${\a \over 3}$}
 \Text(-105,-10)[c]{$0$}
\Text(-30,-10)[c]{$\m_b$}\Text(-60,-10)[c]{$\m_c$}
 \Text(-150,-10)[c]{$-\m_c$}

\end{picture}

\end{center}
\caption{The horizon position $r_h$ as a function of the black
hole mass $\m$ for the Gauss-Bonnet branch $\ep=+1$ and for
$\a>0$, where the vacua are $AdS_4$.} \label{1Horizon-b}
\end{figure}

\subsubsection{The $AdS_4$ vacua}

Vacua with $\k=-1$, which are $AdS_4$, exist only in the
Gauss-Bonnet branch. In the Einstein branch, for all the cases
satisfying the constraint (\ref{papcon2}), the region  shielded by
the horizon is timelike and therefore not suitable for our
compactification.

In the Gauss-Bonnet branch $\ep=+1$, we have only non-compact
brane world models, all of them for $\a>0$. The solutions satisfy
the criteria (\ref{papcon1})-(\ref{papcon3}) for $-\m_c \leq \m
\leq \m_b \equiv 5 \sqrt{6} \a^{3/2} /9$. The singularity
(\ref{chasing}) exists for $0<\m \leq \m_b$ but is always hidden
behind the horizon. The plots of the horizon position as a
function of $\m$ in the above cases is given in
Fig.\ref{1Horizon-b}.

\subsubsection{The Nariai $dS_4$ vacuum} \label{Nariai}

The degenerate case for $\k=+1$ in the Gauss-Bonnet branch
$\ep=+1$ deserves special consideration. This happens when the mass parameter reaches its maximal value $\m \to
 \m_c$ and so the two horizons become one, $r_- \to r_+ \to \sqrt{|\a|/3}$. The internal  space then
appears to collapse, however one
should look as for the Nariai metric to a different coordinate
system \cite{perry}. So, setting $\xi \equiv r_-/r_+$, we make the coordinate
transformation that consists of an affine transformation \ba r&=& {r_+ \over 2} [(1-\xi)R +1+\xi]~,  \\
\th&=&{2 r_+  \over (1-\xi)}\Th ~.
 \ea
The degenerate case corresponds to the limit $\xi \to 1$ and as
can be seen from (\ref{chasol}), it is the limit that the
four-dimensional part of the metric is rendered unwarped. In these
coordinates $R \in [-1,1]$ and $\Th \in [0,2\pi C)$ with \be c= {2
r_+ \over (1-\xi)} C ~. \ee The affine transformation blows up a
point of the coordinate system $(r,\th)$ to a well behaved
internal space in the $(R,\Th)$ coordinates. Let us now define the
modified potential \be f={4 V \over (1-\xi)^2}~, \ee with the
parameters $\mu$ and $|\a|$ as functions of $r_+$ and $\xi$ given
by \be \m=r_+^3 \xi (1+\xi) \quad , \quad |\a|=r_+^2 {1-\xi^3
\over 1-\xi}~. \ee Then, substituting everything in the metric
(\ref{chasol}) we obtain the blown up metric \be ds^2 =
r_+^2\left(f d\Th^2 + {dR^2 \over f} + {r^2 \over r_+^2}
h_{\m\n}dx^\m dx^\n \right)~. \ee The Nariai limit $\xi \to 1$
gives a non-singular limit for the modified potential $f \to {3
\over 5}(1-R^2)$. Therefore, we obtain a non-vanishing internal
space \be ds^2 = \left({|\a| \over 3} \right) \left[ {3 \over
5}(1-R^2)d\Th^2 + {dR^2 \over {3 \over 5}(1-R^2)} + h_{\m\n}dx^\m
dx^\n \right]~. \ee

Note that, after this coordinate transformation,  the deficit
angle from (\ref{antc}) can be expressed as a function of the
modified potential as \be \b_+=\b_-={1 \over 2} |\de_R f| C = {3
\over 5}C~. \ee Thus, in this limit the deficit angle is
independent of $\a$ and depends only on the periodicity $C$ of the
angular coordinate.

\subsubsection{Zero black hole mass $\m=0$} \label{secm=0}

Let us now suppose that also the black hole mass vanishes $\m=0$.
Then, the potential has the simple cosmological form \be
V(r)=\k+{r^2 \over 2\a} (1+\ep)~,  \ee from where we can see that
the Einstein branch $\ep=-1$ can only give the trivial flat
six-dimensional  solution, for $\k=+1$.  The Gauss-Bonnet branch
$\ep=+1$ on the other hand has non-trivial solutions. The flat
$\k=0$ case is excluded since it has no horizon, but there are
$AdS_4$ and $dS_4$ solutions. Firstly, the $\k=-1$ case can have
brane world solutions as long as $\a>0$. Then the spacelike region
that we use for the internal space is non-compact. This case
corresponds to the point of Fig.\ref{1Horizon-b} where the curve
intersects the $\m=0$ axis.

On the other hand, for the $\k=+1$ vacua we can have an horizon
for $\a<0$ and obtain a compact brane world model which is part of a six dimensional de-Sitter space.
This case
corresponds to the point of Fig.\ref{2Horizons} where the upper
curve of $r_+$ intersects the $\m=0$ axis.  Furthermore, since the
space has no singularity at $r=0$, we can extend the radial
coordinate to $r<0$ and consider the region of $-\sqrt{|\a|} \leq
r \leq \sqrt{|\a|}$. The internal space is symmetric around $r=0$,
thus we have $Z_2$ symmetry around the
equator of the internal space. The reintroduction of a black hole
mass $\m \neq 0$ breaks this symmetry since it introduces an $r=0$
singularity.

\subsubsection{Flat vacuum} \label{flat}

As noted before, the black hole ansatz (\ref{chasol}) does not
admit flat vacua for $a^2=0$. However, this ansatz, as we noticed
also in the Nariai vacuum, has limitations, since it describes
only  warped four-dimensional metrics. Looking for unwarped
solutions, it is easy to see that the following flat vacuum is a
trivial solution of the equations of motion (\ref{papein}) \be
ds^2= \e_{\m\n}dx^\m dx^\n + d\r^2 + \r^2 d\th^2~, \ee where $\th
\in [0,2\p c)$. It is non-compact and exists for any sign of $\a$.
The equations (\ref{antNewt}), (\ref{antJunct}),
(\ref{antcontrib}) continue to hold for $\k=0$. To compactify such
flat vacua, a gauge field flux has to be added along the lines of
Appendix A.

\subsection{Non-zero bulk cosmological constant}

Let us now switch on the bulk cosmological constant, \ie $a^2 \neq
 0$. The polynomial (\ref{papcon3}) is more complicated to solve,
therefore we will study some special cases and focus on as few
parameters as possible. Namely, the case with zero black hole mass
$\m=0$ and also the point $1+4\a a^2=0$ in parameter space where
the theory can be written in a   BI form. Finally, we will provide
all the flat ($\k=0$) vacua.

\subsubsection{Zero black hole mass $\m=0$}

If the black hole mass vanishes $\m=0$,   the potential has the
simple cosmological form \be V(r)=\k+{r^2 \over 2\a} [1+\ep
\sqrt{1+4\a a^2}]~. \ee It is obvious that the only brane world
solutions that we can obtain in the present case are the curved
ones $\k=\pm 1$.

\begin{figure}[t]
\begin{center}
\epsfig{file=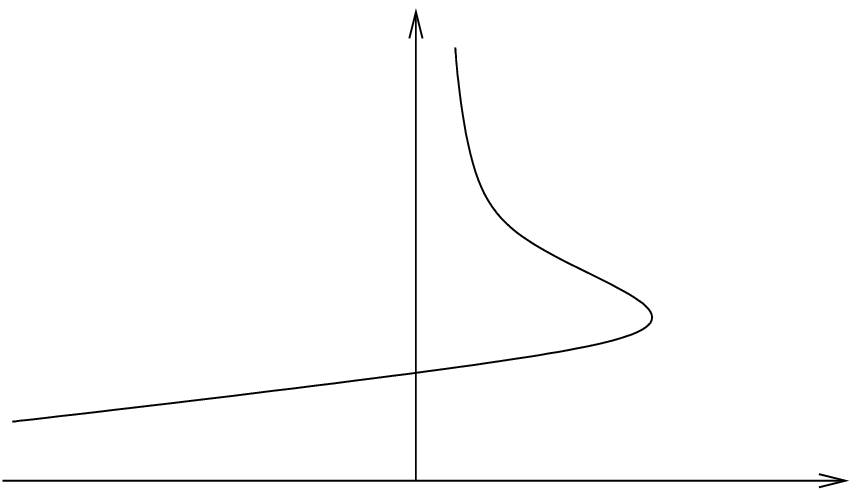,width=5.5cm,height=3.5cm} \hspace{.5cm}
\epsfig{file=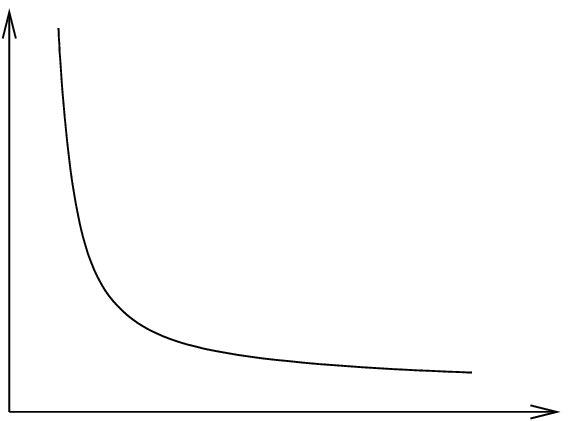,width=4.5cm,height=3.5cm}
\begin{picture}(50,50)(0,0)
\Text(-245,100)[c]{$r_h^2$} \Text(-140,100)[c]{$r_h^2$}
\Text(0,10)[c]{$a^2$} \Text(-150,10)[c]{$a^2$}
\DashLine(-233,36)(-190,36){3} \DashLine(-190,36)(-190,3){3}
\Vertex(-190,36){2} \Text(-190,-10)[c]{$1/(4|\a|)$}
\Text(-130,-10)[c]{$0$} \Text(-233,-10)[c]{$0$}
 \Text(-250,36)[c]{$2|\a|$}
 \Text(-280,60)[c]{$\a<0$}\Text(-212,18)[c]{$\ep=+1$} \Text(-190,58)[c]{$\ep=-1$}
\Text(-60,50)[c]{$\a>0$} \Text(-60,35)[c]{$\ep=-1$}
\end{picture}
\vspace{-.5cm}

\end{center}
\caption{The horizon position $r_h$ as a function of the bulk
cosmological constant $a^2$ for the $dS_4$ vacua with $\m=0$. The
solution on the left is for $\a<0$, where the Einstein and Gauss
Bonnet branches meet at the BI point, represented by dot. The
solution on the right is for $\a>0$.} \label{Lambda}
\end{figure}

Let us first discuss the $dS_4$ ($\k=+1$) solutions. For $\a>0$,
we see from (\ref{papcon2})  that only the Einstein branch
$\ep=-1$ has a solution with a horizon. Furthermore, in this case,
in order to have $r_h^2>0$, the bulk cosmological constant should
be positive, \ie $a^2>0$.  For $\a<0$ we can have solutions both
in the Einstein and the Gauss-Bonnet branch. In the Einstein
branch $\ep=-1$ the horizon is $r_h^2>0$ for positive bulk
cosmological constant with $0<a^2<1/(4|\a|)$. On the other hand,
in the Gauss-Bonnet branch $\ep=+1$ there are solutions for any
$a^2 < 1/(4|\a|)$. At the BI limit $a^2=1/(4|\a|)$ the two
branches merge to a unique $dS_4$ solution. In all the above cases
the spacelike region is for $r<r_h$ and the internal spaces are
compact. As in Sec.\ref{secm=0}, we can extend the radial
coordinate to $r<0$ and consider the region of $-r_h \leq r \leq
r_h$, since the space has no singularity at $r=0$. The plots of
the horizon position as a function of $\m$ in the above cases are
given in Fig.\ref{Lambda}.

The $AdS_4$ ($\k=-1$) solutions are the same as above if we
substitute $a^2 \to -a^2$. Thus, the curves in Fig.\ref{Lambda}
should be the mirrors of the $\k=1$ case with respect to the
$a^2=0$ axis. In addition, the spacelike region for these
solutions is for $r>r_h$, thus all the $AdS_4$ vacua are for
non-compact internal spaces.

\subsubsection{The Born-Infeld  vacua}

\begin{figure}[t]
\begin{center}
\epsfig{file=plot2.eps,width=5cm,height=3.5cm} \hspace{.5cm}
\epsfig{file=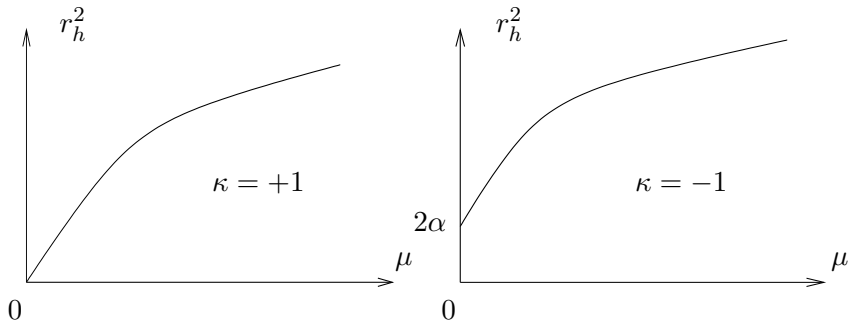,width=5cm,height=3.5cm}
\begin{picture}(50,50)(0,0)
\Text(-290,100)[c]{$r_h^2$} \Text(-125,100)[c]{$r_h^2$}
\Text(0,10)[c]{$\m$} \Text(-165,10)[c]{$\m$}
\Text(-148,-8)[c]{$0$} \Text(-312,-8)[c]{$0$}
 \Text(-155,25)[c]{$2\a$}
 \Text(-220,40)[c]{$\k=+1$}
\Text(-60,40)[c]{$\k=-1$}
\end{picture}
\vspace{-.5cm}

\end{center}
\caption{The horizon position $r_h$ as a function of the black
hole mass $\m$ for the Born-Infeld case $E=-1$ and for $\a>0$. The
left graph is for $dS_4$ vacua and the right for $AdS_4$ vacua.}
\label{1Horizon-BI}
\end{figure}

For the  $dS_4$  and $AdS_4$ vacua, it is easy to see that $V$ is
monotonic for $E=-1$ and has an extremum at
$r_{extr}=(|\a|\m/4)^{1/5}$  if $E=+1$.

The vacua obtained for $1+4\a a^2=0$ deserve special attention. As
discussed in the Sec.\ref{secBH}, these vacua do not have an
Einstein limit. In a certain way they correspond to the strongly
coupled limit of the Gauss-Bonnet term in the action
(\ref{chaaction}), since at the linearized level the combination
$(1+4\a a^2)$ multiplies the perturbation operator. Furthermore,
as we saw in the previous section, for these vacua the branches
for $\ep=+1$, $\ep=-1$ merge. Let us now switch on a mass
parameter $\m>0$ in the potential and write it as \be \label{born}
V(r)=\kappa + \frac{r^2}{2\alpha}+{M\over \sqrt{r}}~, \ee where
the integration constant $M\equiv\frac{E\sqrt{4|\alpha|\m}}{2\a}$~
replaces $\m$. The potential (\ref{born}) is similar to an
Einstein  black hole potential in four dimensions with a
cosmological constant, apart from the fact that
 the Newtonian potential is now $1/\sqrt{r}$ rather than $1/r$! Furthermore,
 if we were to compare it with the six dimensional Einstein black hole
 ($1/r^3$), we see that gravity in the Born-Infeld case is far weaker as one approaches the singularity.

Keeping this comparison in mind, the flat $\k=0$ vacua of this
theory, exist for $E=-1$ and are non-compact. In order for the
region $r>r_h$ to be spacelike, we need $\a>0$ and thus $a^2 <0$
(similar to a planar $AdS$ black hole). The horizon position is
simply $r_h=(4 \a \m)^{1/5}$. We will see more of the flat vacua
in the last subsection.

For the  case $E=-1$, the potential has only one root. In order
for the region $r>r_h$ to be spacelike, we need to have $\a>0$.
Then both cases $\k=\pm1$ are possible. For the $AdS_4$ ($\k=-1$)
vacua, the horizon distance is bounded from below $r_h^2 \geq 2
\a$, while for the $dS_4$ ($\k=+1$) vacua, the horizon  is not
bounded.  The plot of the horizon positions as a function of $\m$
is given in Fig.\ref{1Horizon-BI}. The situation is similar to an
$AdS$ Schwarzschild black hole with planar or spherical horizon.

\begin{figure}[t]
\begin{center}
\epsfig{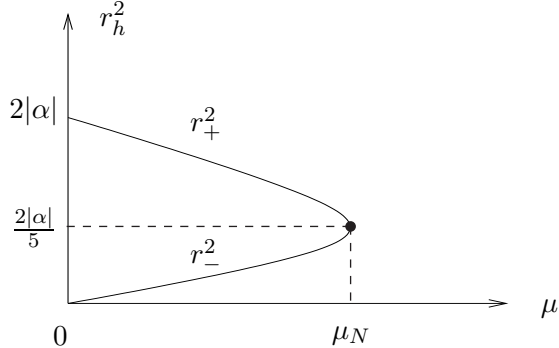}
\begin{picture}(50,50)(0,0)
\Text(-155,110)[c]{$r_h^2$} \Text(10,0)[c]{$\m$}
\DashLine(-172,31)(-65,31){3} \DashLine(-65,31)(-65,3){3}
\Vertex(-65,31){2} \Text(-120,20)[c]{$r_-^2$}
\Text(-120,70)[c]{$r_+^2$} \Text(-65,-10)[c]{$\m_N$}
\Text(-175,-10)[c]{$0$} \Text(-185,31)[c]{${2|\a| \over 5}$}
\Text(-185,75)[c]{$2|\a|$}
\end{picture}

\end{center}
\caption{The horizon positions $r_-$ and $r_+$ as a function of
the black hole mass $\m$ for the Born-Infeld case $E=+1$ and for
$\a<0$, where the vacua are $dS_4$. For $\m=\m_N$ we obtain the
degenerate horizon case. The outer horizon $r_+$ is also the
horizon that we obtain for the case  $E=+1$ and for $\a>0$, where
the vacua are $AdS_4$.} \label{2Horizons-BI}
\end{figure}

In the $E=+1$, the potential has either none or two roots. In
order for the potential to acquire two roots the black hole mass
should be in the range $\m> \m_{N} \equiv 4
(2/5)^{5/2}|\a|^{3/2}$. Then we can distinguish two possible
cases, one with $\a>0$ and $\k=-1$, and a second with $\a<0$ and
$\k=+1$. For the $AdS_4$ ($\k=-1$) vacua, the spacelike region is
for $r>r_+$ and thus the space is non-compact. The horizon is
bounded as $2\a/5<r_+^2<2\a$. For the $dS_4$ ($\k=+1$) vacua, the
spacelike region is for $r_-<r<r_+$ and thus the space is compact.
The outer horizon is bounded as $2\a/5<r_+^2<2\a$ and the inner as
$0<r_-^2<2\a/5$. When $\m=\m_N$ the two roots become degenerate
and $V'_{r_\pm}=0$. This solution will be discussed in the
following section. The plot of the horizon positions as a function
of $\m$ is given in Fig.\ref{2Horizons-BI}.

\subsubsection{The Born-Infeld  Nariai vacuum}  \label{BInariai}

The degenerate case of the Born-Infeld vacuum for $\k=+1$, $E=+1$
and $\a<0$ deserves special consideration. This happens when $\m
\to
 \m_N$ and so $r_- \to r_+ \to \sqrt{2|\a|/5}$. The internal  space then
appears to collapse, however one should look as for the Nariai
metric in a different coordinate system, as we saw in
Sec.\ref{Nariai}. Repeating the same procedure as before and
expressing the parameters $\mu$ and $|\a|$ as functions of $r_+$
and $\xi$  \be \m={r_+^3 \over 2} {\xi (1-\xi^2)^2 \over
(1-\xi^{1/2})(1-\xi^{5/2})} \quad , \quad |\a|={r_+^2 \over 2}
{1-\xi^{5/2} \over 1-\xi^{1/2}}~, \ee we find that the Nariai
limit $\xi \to 1$ gives a non-singular limit for the modified
potential $f \to {1 \over 2}(1-R^2)$. Therefore, we obtain a
non-vanishing internal space \be ds^2 = \left({2|\a| \over 5}
\right) \left[ {1 \over 2}(1-R^2)d\Th^2 + {dR^2 \over {1 \over
2}(1-R^2)} + h_{\m\n}dx^\m dx^\n \right]~. \ee

Furthermore,  the deficit angle from (\ref{antc}) can be expressed
as a function of the modified potential as \be \b_+=\b_-={1 \over
2} |\de_R f| C = {1 \over 2}C ~. \ee

\subsubsection{Flat vacua}

Let us now study the flat ($\k=0$) vacua that exist in the $a^2
\neq 0$ case.  We have found previously a flat vacuum in the
special BI case. More generally, we know from (\ref{papcon2}) that
flat vacua can exist only for $\ep=-1$. The horizon position is at
$r_h^5=-\m/a^2$ and
 we can see from the potential asymptotics that the region $r>r_h$
will be spacelike in two cases: (i) $\a<0$ and $a^2<0$ and (ii)
$\a>0$ and $-1/(4\a)<a^2<0$.  Thus, for both cases we should have
$\m>0$. In case (i) there is a branch cut singularity $r_s$ given
in (\ref{chasing}) hidden always beyond the horizon. In case (ii)
there is not such $r_s$.

In all these cases, the flat vacua have non-compact internal
spaces and are warped. The latter property is in contrast to the
flat vacuum that exists when  $a^2 =0$. Thus, we see that the
introduction of a bulk cosmological constant can be compensated by
the black hole mass $\m$ and warping of the four-dimensional
space, giving again vacua of zero four-dimensional curvature. As
noted before, to compactify such flat vacua, a gauge field flux
has to be added along the lines of Appendix A.

\section{Self-properties of the solutions}

Let us now discuss the physical consequences of the above
solutions. In particular, we wish to see whether we can obtain
codimension-2 braneworlds  exhibiting  {\it self-accelerating} or
{\it self-tuning} behaviour. For the former, we seek $dS_4$ models
where the acceleration is mainly of geometrical origin rather than
due to the vacuum energy or tension $T$ of the branes. For the
latter, we seek braneworld solutions (de Sitter or flat) where the
effective Hubble parameter of the brane is not related to the
brane vacuum energy $T$ and the tension can be arbitrary large
even for small observable Hubble curvature. Here it is important
to emphasize, that the Gauss-Bonnet term plays the role of an
induced gravity term on the brane \cite{zcc} according to
(\ref{antInduced}). Therefore even if the internal space is not
compact we expect that gravity will be four dimensional up to some
distance scale. In fact, we would expect, and this turns out to be
true, that self-acceleration will hold for warped but
uncompactified codimension-2 setups as for the case of
codimension-1. Secondly, taking into account that the effective
curvature of the brane is given by \be \k H_\pm^2 = -{1 \over 2
\a} + {8\pi G_4^\pm \over 3} T^\pm~, \label{antcontrib1}\ee \ie as
the sum of a geometrical term and a brane tension term, the
self-accelerating and the self-tuning cases are, in some respect,
in contrast to each other.

Let us first study {\it self-acceleration}. The key relation is (\ref{antcontrib1}). For this case,
one should firstly have a non-zero positive geometrical
contribution to the curvature, \ie $\a<0$ coming from the Gauss-Bonnet term in the action (\ref{chaaction}).
Secondly, this
contribution should be dominant in comparison to the brane tension
contribution. Note that in order to be in accord with
phenomenology from supernovae data \cite{super} and since $H_0\sim 10^{-34} eV$, the Gauss-Bonnet
coupling should be roughly of the order $\a
\sim 10^{120} M_{Pl}^{-2}$. In this
case, the Gauss-Bonnet term in the action is dominant in
comparison to the Einstein-Hilbert term. Furthermore, for this
limit, even the compact models will have enormous volumes of
cosmological size, and therefore will be comparable to certain aspects of
infinite volume ones (see Appendix C for the calculation of the
volumes of the various models).

It is useful to define a self-acceleration index as the fraction
of the geometrical acceleration to the curvature \be s={1/(2 |\a|)
\over H^2}={r_h^2 \over 2 |\a|}~. \label{saindex}\ee When $0<s <
1$ we have $T>0$ assuming $G_4>0$. On the other hand, if $s>1$, we
have $T<0$ with the limit $|T| \to 3/(16 \pi G_4 |\a|)$ giving $s
\to \infty$. We can argue that self-acceleration exists whenever $
1/2 \lesssim s \lesssim 3/2$, with the strict self-accelerating
limit being $s \to 1$ for $T=0$.

The index (\ref{saindex}) depends on the specific bulk solutions
that we have. Going through the $dS_4$ vacua that we discussed in
the previous section, we should restrict ourselves to the ones
with $\a<0$. Then the self-acceleration index for the various cases is
listed as following

\begin{enumerate}

\item For $a^2=0$, $\ep=+1$, $\a<0$, we have two horizons and the
index for the two branes varies as $0 < s_- < 1/6$ and $1/6 < s_+
< 1/2$. Thus, for this case (Fig.\ref{2Horizons}) we have no
strict self-accelerating limit. The Nariai limit case gives
$s=1/6$.

\item For $a^2=0$, $\ep=-1$, $\a<0$, we have one horizon and the
index is $s > 1$  (Fig.\ref{1Horizon-a}). Here we have a strict
self-accelerating limit when $\m \to \m_s$, however, at $\m=\m_s$
the model is singular.

\item For $a^2=0$, $\ep=+1$, $\m=0$, $\a<0$, we have the compact
$Z_2$-symmetric model with $s=1/2$.

\item For $a^2 \neq 0$, $\ep=\pm 1$, $\m=0$, $\a<0$, we have the
compact $Z_2$-symmetric model (Fig.\ref{Lambda}). The index for
$\ep=-1$ is $s>1$ and for $\ep=+1$ varies as $0<s<1$. The  strict
self-accelerating limit happens for the BI case and can be approached by either
branch $\epsilon=1$ and $\epsilon=-1$. One can argue
that the BI limit is natural in the sense that the theory becomes
more symmetric at that point.

\item For the BI case with $E=+1$, $\m \neq 0$, $\a<0$, we have
two horizons (Fig.\ref{2Horizons-BI}) and the index for the two
branes varies as $0 < s_- < 1/5$ and $1/5 < s_+ < 1$. The strict
self-accelerating limit happens for the $\m=0$ case as mentioned
above.

\end{enumerate}

From the above we see that the two physically interesting cases
where the late acceleration of our Universe can be of purely
geometrical origin, are cases 2 and 4. In case 2 the strict $s=1$
limit can only be reached asymptotically in order not to hit the
singularity at $\m=\m_s$.

Actually, the first two  cases present some similarities and one
important difference with the codimension-1 DGP \cite{dgp}, which
is worth noting. In fact, we see that case 2 corresponds to the
self-accelerating  branch of DGP \cite{deffayet} and case 1 to
that of the normal branch with $dS_4$ branes \cite{kal2}. A first
similarity is the fact that we obtain branching, which now is
in-between the Gauss-Bonnet branch and the Einstein branch.
Secondly, for case 1, as for the normal branch with $dS_4$ branes,
we expect a normalisable spin-2 graviton since the relevant volume
element is finite. It is an intriguing difference, however, that
the codimension-1  unstable or self-accelerating branch
corresponds to the Einstein branch here and not the Gauss-Bonnet
one.

In case 4, we note that the limit $s=1$ is regular and we see that
around $a^2 = 1/(2|\a|)$, self-acceleration is possible for both
branches. Here, the similarity with codimension-1 resides that in
that the case of $T=0$ tension we have enhanced symmetry
\cite{kal,kal2}, as one also expects for the Born-Infeld case in
codimension-2.

Let us now discuss the cases which have to do
with {\it self-tuning}. The self-tuning idea aims to allow for
vacua where their  effective cosmological constant is  independent
from (or at least not strongly dependent on) the vacuum energy of the brane, without fine-tuning. In the
toy-model we are presenting here, vacuum energy is represented by
the brane tension $T_{\pm}$ whereas the effective cosmological
constant is represented by $\k H_\pm$. For self-tuning to operate,
there should be enough integration constants allowed to be
adjusted when we vary one brane tension, with the crucial demand
of keeping the curvature of that brane constant. In the solutions
that we have discussed, the free parameters are the angular
coordinate range $c$ and the black hole mass $\m$. On the other
hand, the action parameters are the Gauss-Bonnet coupling $\a$,
the bulk cosmological constant $a^2$ and the brane tensions
$T_\pm$. If we vary for instance $T_+$ we do not wish $\a$, $a^2$,
$T_-$, $H_+$ to change, but only possibly $c$ and $\m$.  We will
be obviously interested in the $dS_4$ and the flat vacua to test
if such a self-tuning can work.

For the $dS_4$ vacua, the Hubble parameter on the brane depends on
the black hole mass $\m$ and the action parameters $\a$ and $a^2$.
Since we require that the action parameters are not tuned, if we
fix the curvature $H^2$ this is equivalent to fixing the black
hole mass $\m$. In the cases where we have more than one brane
with different Hubble parameters $H_\pm$, both of the latter are
given functions of $\a$, $a^2$ and $\m$. Thus, fixing $\m$ from
the curvature of the one brane $H_+$, fixes also the curvature of
the second brane $H_-$. Then from the relation \be 2\pi
(1-\beta_\pm)\left(\frac{1}{2\alpha}+ \k H_{\pm}^2\right)=\frac{4
\pi G_6}{\alpha} T^{\pm}~, \ee we see that, if we change $T^+$, we
can keep the curvature $H_+$ constant by changing $c$. But since
$H_-$ is fixed, we have to change also $T^-$. This results to an
interbrane fine-tuning. Hence, the only way to obtain self-tuning
is when  the second brane is absent, or in $Z_2$-symmetric models
where only one of the above relations survive.

These selftuning cases, however, are not all theoretically
satisfying if we look at the orders of magnitude of the various
dimensionful quantities in  (\ref{antcontrib1}). Although we have
no idea of the brane tension value today, it would be logical
that,  if no accurate cancellations happen during phase
transitions in the history of the Universe, its natural order of
magnitude is $T \sim M_{Pl}^4$. On the other hand we know that the
the present value of the Hubble constant is approximately $H_0
\sim 10^{-60}M_{Pl}$. In order that (\ref{antcontrib1}) can then
be satisfied, today's curvature should be negligible in comparison
with $|\a|^{-1}$. In other words, we should find models which
allow for $|\a| H^2 \ll 1$ and approximately \be {8\pi G_4 \over
3} T \approx {1 \over 2 \a}~. \label{selfcond} \ee The above
relation should not be viewed as a fine tuning between the action
parameters $\a$ and $T$, since $G_4$ scales as $1 \over \a$
(\ref{antNewt}) and {\it can vary in time} when the vacuum energy
changes, in accordance with the remarks of \cite{nillpaptan}. The
constraints on the time variation of Newton's constant come rather
late in the history of the Universe, certainly after the QCD phase
transition. Therefore, the relation (\ref{selfcond}) is providing
the angle deficit/excess parameter $\b$, given a  Gauss-Bonnet
coupling $\a$, which needs not to be unnatural as in the
self-accelerating case. Furthermore, the relation (\ref{selfcond})
is exact for the case of flat $\k=0$ vacua.

From the models that we have discussed, let us see the cases
that give rise to such theoretically viable self-tuning

\begin{itemize}

\item For $a^2=0$, $\ep=-1$, and  for both  $\a>0$ and $\a<0$, we
have the non-compact vacua with $|\a| H^2 \ll 1$ for $\m
|\a|^{-3/2} \gg 1$ (see Fig.\ref{1Horizon-a}).

\item For $a^2 \neq 0$, $\ep=-1$, $\m=0$, $\a<0$, we have
$Z_2$-symmetric vacua with $|\a| H^2 \ll 1$ for $|\a| a^2 \ll 1$
(see Fig.\ref{Lambda}).

\item  For $a^2 \neq 0$, $\ep=-1$, $\m=0$, $\a>0$, we have
non-compact vacua with $\a H^2 \ll 1$ for $\a a^2 \ll 1$ (see
Fig.\ref{Lambda}).

\item For the BI case with $E=-1$, $\a>0$, we have non-compact
with $\a H^2 \ll 1$ for $\m \a^{-3/2} \gg 1$ (see
Fig.\ref{1Horizon-BI}).

\item All the flat non-compact vacua, for both $\a>0$ and $\a<0$
are self-tuning in the exact sense.

\end{itemize}

Here we should note once more that the above (exact or approximate) self-tuning
solutions do not constitute a resolution, but rather a potential amelioration, of the
cosmological constant problem. We have found vacua where the brane curvature is
independent of the brane vacuum energy and furthermore the these two energy scales
can be well separated. However, the important question, for which the self-tuning
mechanism cannot give a simple answer, is if the dynamical variation of the vacuum energy
results in remaining in these vacua. In some sense, the term self-tuning itself is
deceiving since the existence of these vacua does not guarantee that there are some attractor
dynamics which tune the system towards them. Nevertheless, the very existence of them
is important, since at least at the level of non-dynamical solutions, dissociating the brane
curvature from its vacuum energy is a rather unique situation. Moreover, we saw that most of 
these vacua are non-compact, but, as we will explain in the last section, gravity is quasi-localised 
and thus effectively four dimensional up to some cross-over scale.

It is evident that the vacua with the above self-tuning properties
are different from the ones with the self-accelerating properties
corresponding in particular to a totally different bare parameter
$\alpha$.  This is true as long as we want self-acceleration at a
very low energy scale. If, for example, we want to explain
inflation theory geometrically, as in \cite{staro}, then the
self-accelerating (inflationary) vacua can be deformed to
self-tuning ones by letting the integration constant $\mu$ run. A
nice toy model scenario for this is pictured in
Fig.\ref{1Horizon-a} if we assume that $\mu$ is an increasing
function of proper cosmological time. One starts at the big-bang
singularity which corresponds to a 6 dimensional bulk  naked
singularity at $\mu=\mu_s$. At early time we have  $\mu\gtrapprox
\mu_s$ and the inflationary expansion is, $H_{inf}^2 \sim
1/(2|\alpha|)$. At late time the late expansion is related to the
large mass of the soliton whereas the vacuum energy is of the
order of $1/(G_4|\a|)$ which is much larger than the current
Universe curvature $H_0^2 M_{Pl}^2$.

Before we close off this section, it is worth pointing out that no
self-tuning cases are possible for the Gauss-Bonnet branch.
Furthermore, the only finite volume self-tuning case that we
encountered is possible for a de Sitter braneworld.

\section{The Gauss-Bonnet instantons}

In our previous analysis, we have found several brane world vacua
of finite volume. These solutions are of special interest because
 they can give rise to gravitational instantons. We saw how they all have the feature
that the four dimensional space is $dS_4$. Wick rotating to a
Euclidean metric $h_{\m\n}$ as in (\ref{paph}) with $\k=+1$ and
keeping the Euclidean internal space, all the coordinates are
spacelike and the solutions describe a Riemannian manifold with
conical singularities. Particularly important are the instantons
which are regular, in other worlds those having no conical
singularities (see also \cite{newper,gbins}).
 We can divide these vacua into two classes according to their
topology.

The first class of instantons are the ones for vanishing black
hole mass $\m=0$, where there is no singularity at $r=0$ and the
space is $Z_2$-symmetric. These instantons are given by \be
ds^2=\left(1-{r^2 \over r_h^2(\a,a^2)}\right)
d\th^2+\frac{dr^2}{\left(1-{r^2 \over r_h^2(\a,a^2)}\right)}+r^2
h_{\m\n} dx^\m dx^\n~, \ee with $r_h^2 = -2\a/(1+\ep\sqrt{1+4 \a
a^2})$ and are topologically $S^6$. In order for these instantons
to be regular, one should have $\b=1$, which from (\ref{antc})
gives $c=r_h$ for the periodicity of the $\th$-coordinate. These
instantons have an Einstein theory limit only in the case $\ep=-1$
and $1+4 \a a^2 \neq 0$.

The second class of instantons are the ones of the Nariai limits
of vacua with non-vanishing black hole mass found in
Sec.\ref{Nariai} and Sec.\ref{BInariai}. These instantons are
given by \be ds^2= r_h^2(\a,a^2) \left[ \z(1-R^2)d\Th^2 + {dR^2
\over \z (1-R^2)} + h_{\m\n}dx^\m dx^\n \right]~,
\label{funnyinst}\ee where $\z$ is a numerical factor and $r_h$
the degenerate horizon. These instantons, in contrast with the
ones noted before, are topologically $S^2 \times S^4$. In order
for these to be regular, one should have $\b=1$. The deficit angle
for the instantons that we have found is given by $\b=\z C$, thus,
regularity of the instantons imposes $C=1/\z$. Let us also note
here that these $S^2 \times S^4$ instantons for the case $a^2 \neq
0$, exist also for the case other that the BI limit that we have
discussed in Sec.\ref{BInariai}. This is why in (\ref{funnyinst})
we allowed for $a^2$ dependence of the degenerate horizon. None of
these instantons has an Einstein theory limit.

The above instantons have the physical interpretation of
describing probabilities of nucleation processes
\cite{perry,lemos} between two distinct gravitational backgrounds.
In particular,
 the probability of nucleation  of a pair of Nariai black holes
from the pure $dS_6$ backgrounds of $\m=0$, is given by \be \G=
\eta \exp[2 \D S] \quad , \quad \D S \equiv S_{S^6}-S_{S^2 \times
S^4} ~, \ee where $\eta$ is the one loop contribution from the
quantum quadratic fields and $S_{S^6}$ and $S_{S^2 \times S^4}$
are the values of the action for the two instantons that we
presented earlier. It is straightforward to compute the above
probability of nucleation for
 the two simple cases of $a^2=0$ and the BI case  $a^2=1/(4
|\a|)$. In both cases one obtains \be \D S \propto  {{\rm
Vol}(S^4) \over 16 \pi G_6 } \a^2 >0 ~. \ee Since these action
differences are positive definite, we deduce that the probability
of both processes is  unsuppressed unlike the case of
\cite{perry}. Therefore, the solitonic vacua ($\m \neq 0$) are
apparently stable with respect to pure de Sitter vacua in Lovelock
theory. This result certainly deserves further study.

\section{Discussion and Conclusions}

In this paper we  studied in some detail codimension-2 braneworlds
in the context of Lovelock gravity. Using a modified version of
Birkhoff's staticity theorem \cite{birk,zegers}, we found all six
dimensional  solutions describing a de Sitter, flat or anti de
Sitter braneworld of codimension 2. First thing we can observe is
that, unlike the codimension-1 case where the Gauss-Bonnet
invariant plays a somewhat secondary role, here, in the case of
codimension-2 braneworlds, it can give rise to self-acceleration
and certain self-tuning properties which are not present in the
Einstein theory. This is largely due to the fact that
codimension-2 junction conditions induce on the brane an
Einstein-Hilbert term \cite{bostock}. It is important to note that
this does not mean that we have ordinary four-dimensional gravity,
rather, as noted in \cite{zcc}, we are in a quite similar
situation as for the codimension-1 DGP model \cite{dgp}. By this,
we mean that from some ultraviolet scale up to some infrared scale
we expect gravity to "look" four-dimensional. A proof of this
statement of course requires the full spectrum of gravitational
perturbations for the solutions that we have found.

The important relation we came across is (\ref{antcontrib}) which
relates the vacuum energy or tension $T$ of the brane with the
Gauss-Bonnet coupling $\a$ and the effective cosmological
expansion on the brane $H_0$, which itself is related to the
characteristics of the bulk solution: its mass, bulk cosmological
constant  (and charge) in particular.  The interesting feature we
found here is that the Gauss-Bonnet coupling can give de Sitter
branes without vacuum energy on the brane, purely geometrically.
In this sense such de Sitter solutions are self-accelerating. In
order to explain the small cosmological constant we then have to
fine-tune as usual $\a \sim H^{-2}_0$.

The second important point is that Gauss-Bonnet coupling $\a$ and
the topological parameter $\beta$, which is otherwise
unconstrained, dictate the cross-over scale between the
four-dimensional Planck scale $G_4$ and the six-dimensional one
$G_6$ (\ref{antNewt}). Here, unlike the five dimensional DGP
model, the cross over scale is not necessarily tied up to the
self-acceleration scale. Indeed, the more $\beta$ is close to 1
(\ref{antNewt}), the more we can dissociate these scales. In a
nutshell, all depends on the gravity phenomenology beyond the
cross-over scale and is dictated by the full graviton propagator
on the brane. As we emphasized, the appearance of the induced
Einstein tensor in the codimension-2 junction conditions is not a
proof of an effective four-dimensional gravity or not. It is a
definite sign, however, that a relevant scale will appear in the
boundary conditions for the gravitational perturbations. Although
we have not undertook the precise perturbation calculation here,
we will comment on some of its characteristics later on.

The most conservative approach is, as in DGP \cite{deffayet}, to
introduce a hierarchy of scales between $G_6$ and $G_4$ so that
the cross-over scale is \be r_c^2={ G_6 \over G_4} ~.
\label{cross} \ee Assuming that $r_c \sim H_0^{-1}$, this gives a
very low higher dimensional Planck scale  $M_6 \equiv G_6^{-1/4}
\sim 10^{-30} M_{Pl}$ which in turn dictates that $1-\beta\sim
{\cal O}(1)$. But, as mentioned above, one can have $r_c \ll
H_0^{-1}$ by having $1-\b \approx 0$. For the self-tuning vacua,
on the other hand, the natural value of the Gauss-Bonnet coupling
is $\a \sim M_{Pl}^{-2}$ which for the same cross-over scale
(\ref{cross}) and $r_c \sim H_0^{-1}$ gives a huge angle excess
$1-\beta\sim 10^{120}$. Obviously, in this case a reasonable
hierarchy between $H_0^{-1}$ and the cross-over scale $r_c$,
cannot reduce significantly the latter huge excess angle. We
emphasize, however, that the above orders of magnitude, should be
viewed with caution since the definition of the cross-over scale
(\ref{cross}) should be done in a true cosmological setting with
complete knowledge of the modified  FRW equations.

Another important  point concerns the fact that the
self-accelerating branch of DGP seems to be embedded in the usual
Einstein branch of Lovelock theory and not in the Gauss-Bonnet
branch. Indeed, given the Gauss-Bonnet term in the action and no
bare cosmological constant ($a^2=0$) we find two distinct
solutions for $\epsilon=\pm 1$ (\ref{chapot}): in the Gauss-Bonnet
branch, $\epsilon=+1$, we can have a finite volume soliton
solution since then the Gauss-Bonnet coupling plays the role of a
cosmological constant (Fig.\ref{2Horizons}) (situation akin to a
Schwarzschild de Sitter  black hole \cite{ms}). Then, we have a
finite volume element and we expect a normalisable zero mode
graviton. We get no self-acceleration. For the Einstein branch
however, $\ep=-1$, we have a single brane, infinite volume element
and self acceleration with as small tension as we want
(Fig.\ref{1Horizon-a}). The bulk solution corresponds to a
Gauss-Bonnet corrected Schwarzschild black hole. This difference
may be interesting in respect to linear (in)stability and the
presence of ghosts of such backgrounds, not only in the scalar,
 but also in the spin-2 sector.

Furthermore, in our solutions we see at least two different length
scales emerging where we can probe differing four-dimensional
gravity. That of the volume element, which if finite, would mean
that beyond that scale we expect ordinary four-dimensional
gravity. Secondly, that of the cross-over scale where up to that
distance gravity seems four-dimensional, as in the case of the
codimension-1 DGP model. Again we emphasize that the cross over
scale and self acceleration scale are here supplemented by an
extra topological parameter $\beta$ which in a sense tells us how
far we are from a Kaluza-Klein setup for the Killing direction
$\partial_\theta$. Indeed, we see that for $\beta \to 0$ our
codimension-2 space is very much like a lightning rod \cite{kk},
where we seemingly have a cascade from a six-dimensional to a
five-dimensional and then a four-dimensional setup (reminiscent of
the recent proposal \cite{cascade}).

What is missing from our analysis  is the linear perturbation of
these solutions which will tell us of the stability and the
precise gravitational spectrum. This is not a trivial task, for
our metric is not conformally flat. Therefore, apart from the
usual complications of black hole perturbations, one has to add
the fact that Lovelock perturbations are going to be genuinely
different from the usual ones for Einstein theory. The reason is
simple: the background Weyl tensor appears in the Lovelock field
equations and thus, extra tensor pieces are bound to appear for
tensorial perturbations. Recent work in fact from string theory
\cite{shenker}, in relation to the $AdS$/CFT correspondence,
indicates that there is  yet another scale appearing in the bulk
perturbations of (\ref{chabh}). This extra scale comes from the
extra tensorial pieces "to be added" to the usual perturbation
operator, that dictate that gravity waves in a planar ($\kappa=0$)
black hole background (\ref{chabh}) can evolve in a differing
background geometry than that of (\ref{chabh}). This means in
particular that the light-cone of the wave fronts can break
causality.  We think that this is an intriguing and extremely
important issue, which we hope will be undertaken soon.

In addition, the cosmology of codimension-2 branes is very poorly
understood, especially from the point of view of self-tuning and
the issue of vacuum selection. It has been known that in Einstein
gravity the introduction of matter other than tension on
codimension-2 branes introduces singularities\footnote{This is a
well known fact in four dimensional gravity, called the Israel
paradox concerning self-gravitating cosmic string metrics (see for
example \cite{paradox}).} worse than conical \cite{Cline} (see,
however, the work of \cite{inter} for intersecting brane cosmology
in six dimensions). Although, the appearance of singularities is
natural in defects of codimension higher than one
\cite{roberto,kalkil}, one has in practice to regularise the brane
\cite{regular,ring} in order to consider some cosmological fluid
on them \cite{ringco,regbraneco}. Although one expects that brane
singularities persist in the Lovelock theory, the fact that there
exists an induced Einstein equation on the brane, allows for the
cosmology to be studied without the need of explicit
regularisation. We plan to address the question of cosmology of
the present Lovelock models in a different publication
\cite{kofi}.

\section*{Acknowledgments}

It is a great pleasure to acknowledge discussions and numerous insightful comments from  Nemanja D. Kaloper, Kozuya Koyama and Antonio H. Padilla. We thank them also for reading through the manuscript. We also thank Roy
Maartens for bringing up the issue of self acceleration in the
context of codimension-2 braneworlds early on. C.C. also thanks Emilian Dudas and Valery Rubakov for discussions and critical comments. A.P. is supported
by a Marie Curie Intra-European Fellowship EIF-039189.

\def\theequation{A.\arabic{equation}}
\setcounter{equation}{0}
\vskip0.8cm \noindent {\Large \bf Appendix A: A more general
action} \vskip0.4cm \noindent

In this Appendix we will consider a generalization of the model
that we discussed in the main text to a $D$-dimensional bulk
spacetime, where also  a gauge field $F_{MN}$  is coupled to
gravity.  The action of the system reads \be S=\int d^D x\,
\sqrt{-g}\left[{1\over 16 \pi G_D}(R +\hat{\alpha}
\L_{GB})-\La\right] -\quarter \int d^D x\, \sqrt{-g} F^2~ , \ee
where  $G_D$ the $D$-dimensional Newton's constant.

It is straightforward to write down the Einstein equations of
motion for the above action. They read \be G_{MN} -
\hat{\a}H_{MN}=8 \pi G_D T_{MN}~, \ee with the energy momentum
tensor
 \be T_{MN}=-\La
g_{MN} + F_{MK} F_N^{~K} - {1 \over 4} F^2 g_{MN}~. \ee
Furthermore, the gauge field equation reads \be \de_M (\sqrt{-g}
F^{MN})=0~. \ee

The static spherically symmetric solutions of the above equations
of motion with $(D-2)$-dimensional space of maximal symmetry are
locally isometric to \be ds^2=-V(r) dt^2+\frac{dr^2}{V(r)}+r^2
h_{\mu\nu} dx^\mu dx^\nu~. \ee The explicit solution of the
equations of motion then gives
 \be F=\frac{q}{4\pi r^{D-2}}
dt\wedge dr ~, \ee and for the potential, \be  V(r)=\kappa +
\frac{r^2}{2\alpha}\left[1+\epsilon \sqrt{1+4\alpha
\left(a^2-\frac{\ep
\mu}{r^{D-1}}-\frac{Q^2}{r^{2(D-2)}}\right)}\right]~, \ee and the
parameters appearing in the action have been rescaled to, \be
\alpha=(D-3)(D-4)\hat{\alpha}~,\qquad 16 \pi G_D
\Lambda=(D-1)(D-2)a^2~\mbox{(for}~ dS_D)~. \qquad \ee

The integration constants are, \be  Q^2=\frac{G_D~ q^2}{2\pi
(D-2)(D-3) \a}~, \qquad \mu=\frac{16\pi G_D M}{(D-2)
\Sigma_{\kappa}}~, \ee where $q$ is the charge, $M$ is the AD or
ADM mass of the solution and $\Sigma_{\kappa}$ is the area of the
unit $(D-2)$ maximally symmetric subspace. Finally,
$\epsilon=\pm1$ which  gives rise to 2 branches of solutions. The
convention of the $\mu$ sign is chosen so that the gravitational
mass is always $\mu>0$. As one can easily check by expanding the
square root for large distances, the sign flip in front of $\mu$
is necessary to match the Schwarzschild solution behaviour for
positive AD mass. On the other hand, the charge term of the above
potential is the opposite of a charged black hole for $\ep=-1$.

\def\theequation{B.\arabic{equation}}
\setcounter{equation}{0}
\vskip0.8cm \noindent {\Large \bf Appendix B: The general solution
of $x^3+ A x + M =0$} \vskip0.4cm \noindent

To determine the positions of the horizons for the $a^2=0$ case,
we need the general solutions of the third order algebraic
equation \be x^3+A x + M =0~, \ee The solutions of the above
equation depend on the parameters $A$, $M$ and in particular on
the discriminant of the system  \be D=\left({M \over
2}\right)^2+\left({A \over 3}\right)^3~, \ee If $D>0$ there is
only one real root. If $D \leq 0$ there are three real roots, two
of which become degenerate where the inequality saturates. A
critical
 value of $M$ that will be useful in the subsequent study is
\be M_c= {2 \over 3 \sqrt{3}} |A|^3~. \ee Since the solutions of
the above equation will correspond to horizons, we also need to
look which of these real roots are in addition {\it positive}. All
permissible cases are listed as following

\begin{figure}[t]
\begin{center}
\epsfig{file=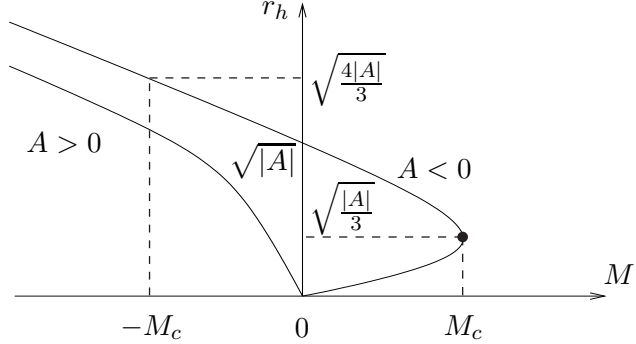,width=8cm,height=4cm}
\begin{picture}(50,50)(0,0)
 \Text(-130,110)[c]{$r_h$} \Text(0,10)[c]{$M$}
 \DashLine(-118,24)(-59,24){3} \DashLine(-59,24)(-59,2){3}
\Vertex(-59,24){2} \DashLine(-177,2)(-177,82){3}
\DashLine(-177,84)(-120,84){3} \Text(-103,84)[c]{$\sqrt{{4
|A|\over 3}}$} \Text(-135,53)[c]{$\sqrt{|A|}$}
\Text(-105,35)[c]{$\sqrt{{|A| \over 3}}$} \Text(-70,50)[c]{$A<0$}
\Text(-210,60)[c]{$A>0$} \Text(-120,-10)[c]{$0$}
\Text(-59,-10)[c]{$M_c$}
 \Text(-177,-10)[c]{$-M_c$}

\end{picture}

\end{center}
\caption{The positive real roots of the equation $x^3+A x + M=0$.
The lower curve gives  the root  for the $A>0$ case and the upper
one the root(s) for the $A<0$ case. In the interval $0<M<M_c$,
there are two roots $r_-<r_+$ which become degenerate for
$M=M_c$.} \label{Genplot}
\end{figure}

\begin{itemize}

\item If $A>0$ and $M<0$, it is $D>0$ and we have the root \be
r_h=\left[-{M \over 2}+ \sqrt{D}\right]^{1/3}-{A \over 3}\left[-{M
\over 2}+ \sqrt{D}\right]^{-1/3}~, \label{papsol1}\ee

\item If $A<0$ and $M<-M_c$, it is $D>0$, but we have one positive
root which is given by the same formula as above (\ref{papsol1}).
[It has different value, however, since $A<0$.]

\item If $A<0$ and $0<M \leq M_c$, it is $D \leq 0$ and we have
two roots \ba r_+&=&2 \sqrt{|A| \over 3} \cos\left( {1 \over 3}
\cos^{-1}\left[
-{M \over M_c}  \right] \right)~, \label{papsol2} \\
r_-&=& \sqrt{|A| \over 3}  \left\{ -\cos\left( {1 \over 3}
\cos^{-1}\left[ -{M \over M_c} \right] \right) \right.  \nn \\
&& \quad \quad \quad \quad \quad \left. +\sqrt{3} ~\sin\left( {1
\over 3} \cos^{-1}\left[ -{M \over M_c}  \right] \right)
\right\}~. \label{papsol3} \ea

\item If $A<0$ and $-M_c \leq M <0$, it is $D \leq 0$ but we only
have one positive root, given by (\ref{papsol2}).

\end{itemize}

The results of the above analysis are summarized in
Fig.\ref{Genplot}.

\def\theequation{C.\arabic{equation}}
\setcounter{equation}{0}
\vskip0.8cm \noindent {\Large \bf Appendix C: Volume calculation}
\vskip0.4cm \noindent

In this Appendix, we will calculate the volume of the general
brane world models that we considered in the main text. To do
this, we will make the assumption that the zero mode graviton
wavefunction follows the warp factor. In other words, we will
assume  that the zero mode is emanating from the metric \be ds^2=
V(r) d\theta^2+\frac{dr^2}{V(r)}+r^2 g^{(4)}_{\m\n}(x) dx^\m dx^\n
~, \label{pertvol} \ee with the four dimensional metric
$g^{(4)}_{\m\n}(x)=h_{\m\n}+H_{\m\n}$, where $H_{\m\n}$ is a
perturbation around the background $h_{\m\n}$. To find the volume
of the space we should substitute the metric (\ref{pertvol}) in
the action (\ref{chaaction}) and integrate out the internal space.
We will define the volume (Vol) to be given by the coefficient of
the four-dimensional curvature term in the action
as\footnote{Notice that according to the ansatz (\ref{pertvol}),
it is $[r]=[\th]=M^{-1}$ and thus $[x^{\m}]=M^0$. For this reason,
we obtain an unusual mass dimension for the volume $[({\rm
Vol})]=M^{-4}$. With a suitable four-dimensional coordinate
rescaling, \eg  $x^\m_N = r_h x^\m$, the volume will be redefined
as $({\rm Vol})_N = ({\rm Vol})/r_h^2$ and will acquire the
correct $M^{-2}$ mass dimension.} \be S = {{\rm (Vol)} \over 16
\pi G_6}\int d^4x \sqrt{-g_4} R_4 + \dots ~, \ee where the dots
represent higher order in curvature terms. It is straightforward
to compute that the various quantities of the action
(\ref{chaaction}) for the ansatz
(\ref{pertvol}) give \ba \sqrt{-g_6}&=& r^4 \sqrt{-g_4} ~, \\
R_6&=& {1 \over r^2} R_4 + \dots ~,
\\ {\cal L}_{GB}&=& {4 \over r^2}\left({V \over r}\right)' R_4 + \dots \ea

Then the volume is given by the following integral \be ({\rm
Vol})= 2\pi c \int dr ~r^2\left[ 1+ 4 \hat{\a} \left({V \over
r}\right)' \right]~. \ee If $1+4\a a^2 \neq 0$, the above integral
gives \be ({\rm Vol})= {2\pi c \over 3} \left[ r^3- {2\a \over
3}\left(4\k r +{5 \m \over \sqrt{1+4\a a^2} }{F \over r^2} \right)
\right]_{r_-}^{r_+} ~, \ee where F is the hypergeometric function
\be F= \!{}_2F_1\left({2 \over 5}, {1 \over 2}, {7 \over 5};{4 \a
\ep \m \over (1+4\a a^2)r^5} \right) ~. \ee

For the special BI case, $1+4\a a^2 =0$, the integral gives
instead \be ({\rm Vol})= {2\pi c \over 3} \left[ {13 \over 3}r^3+
4\a \k r \right]_{r_-}^{r_+}  ~. \ee

It is straightforward to see that the above formulas give finite
(and positive) volume  for the two Nariai cases. To obtain the
volume in this case one should move to the $(R,\Th)$ coordinates
and take the limit $\xi \to 1$.

\end{document}